\pdfoutput=1
\documentclass[twoside]{LCWS11}
\usepackage[latin1]{inputenc}
\usepackage[pdftex]{graphicx,epsfig,color}
\usepackage{wrapfig,rotating}
\usepackage{amssymb,amsmath,array}
\usepackage{cite}
\begin{document}
\title{
ILD Machine-Detector Interface\\ and Experimental Hall Issues} 
\author{Karsten Buesser
\vspace{.3cm}\\
Deutsches Elektronen-Synchrotron DESY \\
Notkestrasse 85, 22607 Hamburg, Germany
}

\maketitle

\begin{abstract} 
The International Large Detector (ILD) is one of the proposed detector concepts for the International Linear Collider (ILC). The work on the ILD machine-detector interface~(MDI) concentrates on the optimisation of the experimental area design for the operation and maintenance of the detector. The ILC will use a push-pull system to allow the operation of two detectors at one interaction region. The special requirements of this system pose technical constraints to the interaction region and the design of both detectors.
\end{abstract}

\section{Introduction}

The Machine-Detector Interface (MDI) work at the International Linear Collider (ILC)~\cite{Brau:2007zza, Elsen:2011zz} covers all aspects that are of common concern to the detectors and the machine. This comprises topics like the mechanical integration of the detectors and the machine as well as aspects of beam induced backgrounds, common instrumentation for beam diagnostics (polarisation, beam energy) as well as common services. Recently, the collaborative work between both detector concepts, ILD~\cite{Group:2010eu} and SiD~\cite{Aihara:2009ad}, together with the ILC machine groups concentrates on the definition and the engineering design of the infrastructures in the experimental area.

\section{The ILC push-pull scheme}

Other as in a storage ring, the total integrated luminosity in a linear collider does not scale with the number of interaction regions. The violent beam-beam interaction degrades the beam quality after the collisions so far, that each bunch of particles can only be used once and is disposed in the beam dump afterwards. Nevertheless, it is a broad consensus, that two complementary detectors, that are run by two independent collaborations, are mandatory to exploit the benefits of healthy competition and to allow for independent cross-checks of the measurements. While the beam delivery system of the ILC is a complex and rather expensive system, a duplication of the interaction region is excluded for economic reasons. Therefore the ILC design foresees to have two detectors that share one interaction region in a push-pull operation scheme. In that scheme, one detector would take data, while the other one is waiting in the close-by maintenance position. At regular schedules, the data-taking detector is pushed laterally out of the interaction region, while the other detector is being pulled in. As the data taking intervals for each experiment should be short enough to avoid a potential discovery by one detector alone, the transition time for the exchange of the detectors needs to be short, i.e.\ in the order of one day, to keep the total integrated luminosity at the ILC high.

The technical design of the push-pull system is a novel and challenging engineering task. Of importance is the definition of the interfaces and boundary conditions that are needed to allow for the friendly co-existence of the two experiments and the accelerator. The top level interfaces and requirements have been laid down in a common publication of the detector concepts and the ILC machine group~\cite{Parker:2009zz}. Major requirements that needed to be defined comprise among others geometrical boundary conditions, vibration tolerances, alignment requirements, vacuum conditions, radiation and magnetic environment.

\begin{figure}[t]
\centerline{\includegraphics[width=0.405\columnwidth]{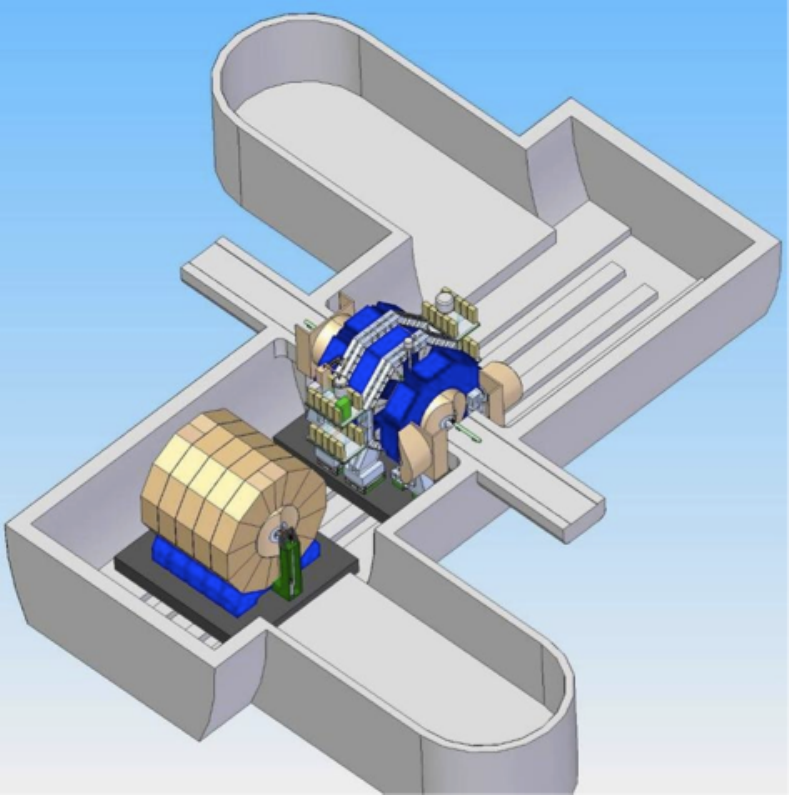}
\includegraphics[width=0.6\columnwidth]{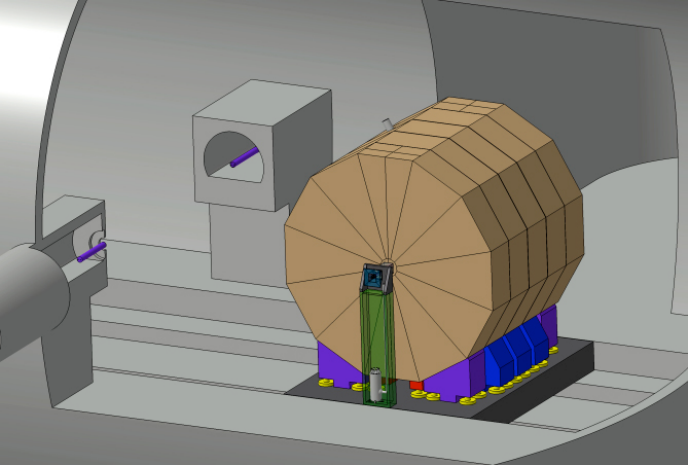}}

\caption{The push-pull system at the ILC: SiD and ILD in the experimental hall (left); ILD on the transport platform (right) \cite{Oriunno:2011}.}\label{Fig:Push-pull}
\end{figure}

One important subject was the conceptual development of the detector transport system for the push-pull operations. The recently agreed upon scheme foresees a platform-based movement system, where each detector would be placed on a platform of reinforced concrete that runs on a suitable transportation system, e.g.\ a rail based system or on air pads. Of major concern were the possible amplifications of ground motion in this system; detailed simulations that were cross-checked with measurements at existing structures showed that these effects can be controlled~\cite{Oriunno:2011}. Therefore the platform system has been adopted as the baseline for the experimental area design. Figure~\ref{Fig:Push-pull} shows a possible configuration of the ILC experimental hall with both detectors on a platform-based push-pull system.

\section{The ILD detector}
The International Large Detector (ILD) is one of the two detector concepts that are under study for the ILC. The design follows a multi-purpose high energy physics detector design and is optimised for the high precision measurements that are anticipated for the ILC~\cite{Group:2010eu}. The size of ILD is roughly 16~m (width) $\times$ 16~m (height) $\times$ 14~m (length); it has a mass of $\approx$~15.5~kt.

\subsection{Assembly and maintenance}

The mechanical design of the ILD detector is inspired by the CMS experiment at the LHC. The main parts are the five rings of the iron yoke, three in the barrel part and two end caps. The detector will be pre-assembled and tested in a surface building. The large assembly pieces will then be lowered into the experimental hall through a large vertical access shaft. The dimensions of the shaft and of the (temporary) crane for these operations are given by the masses and dimensions of the biggest assembly piece. In the case of ILD that would be the central yoke ring that carries the solenoid coil. A shaft diameter of $\approx$~18~m and a hoist for $\approx$~3500~t mass is needed for this.

\begin{figure}[t]
\centerline{\includegraphics[width=0.65\columnwidth]{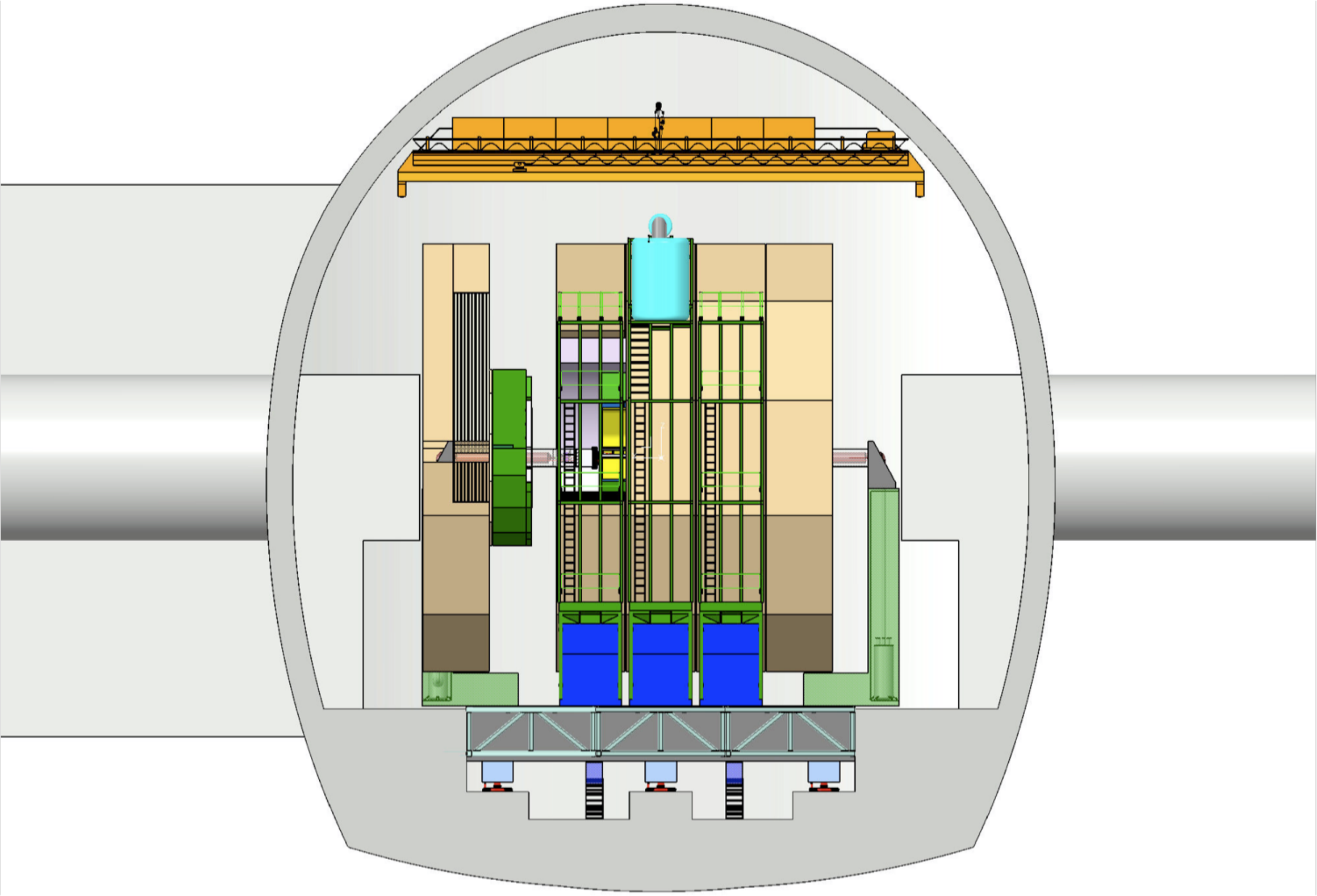}}
\centerline{\includegraphics[width=0.65\columnwidth]{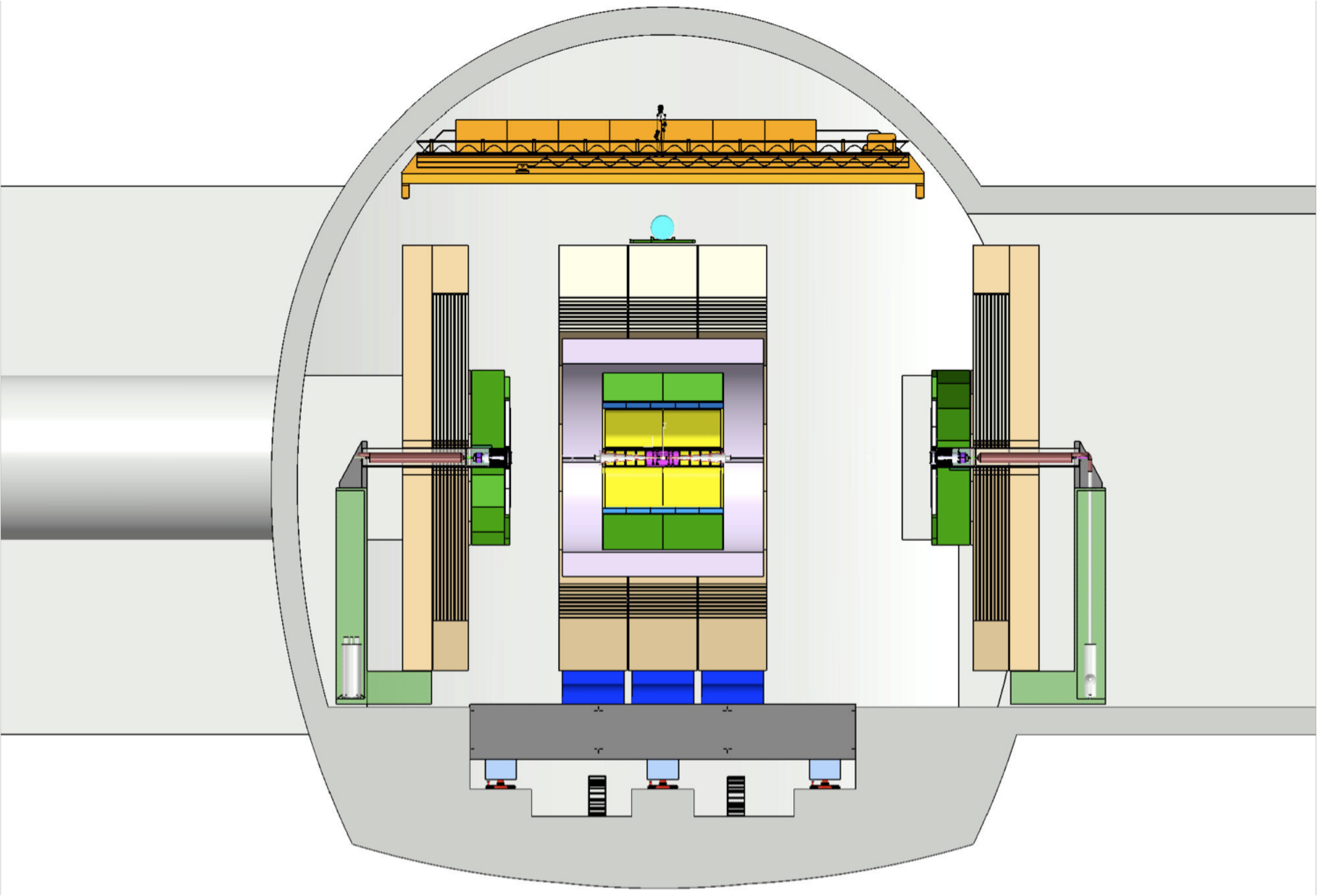}}

\caption{ILD detector opened at the beam line (top) and in the maintenance position (cut view, bottom)~\cite{Group:2010eu}.}\label{Fig:ILD_opening}
\end{figure}

The five yoke rings are mounted on air pads and can therefore be moved easily within the underground experimental hall. In the beam position and during the push-pull movement, the detector is mounted on the transport platform. In the maintenance position, the detector can be opened and the yoke rings can move independently away from the platform. Figure~\ref{Fig:ILD_opening} shows the detector on the beam position and in the maintenance area. The hall layout needs to foresee enough space in the maintenance position to allow the complete opening of the detector rings. During maintenance, access to the inner detector parts and the removal of the large detector components (e.g.\ the time projection chamber) needs to be possible.

\subsubsection{Modified assembly scheme}
\label{Sec:Japan}
Possible ILC sites in Asia (Japan) are different to the other reference sites as they are situated in mountainous regions where a vertical access to the experimental hall might not be given. Instead, horizontal tunnels of $\approx$~1~km length might serve as access ways into the underground experimental area. As the tunnel diameters and the transport capacity within the tunnels are limited for technical and economic reasons, a modified assembly scheme for the ILD detector is under investigation for these sites. In these cases, it is foreseen to still pre-assemble the detector parts on the surface. The yoke rings are however too big and heavy and can therefore be only assembled in the underground hall. The yoke would be transported in segments into the hall where enough space for the yoke assembly and the necessary tools needs to be provided. The largest part of the ILD detector, that should not be divided and therefore needs to be transported in one piece, is the superconducting solenoid coil. Its outer diameter of $\approx$~8.7~m puts stringent lower limits on the diameter of the access tunnel. A considerable effort has been started recently to define the requirements to the detector assembly for the mountain sites.

\subsection{Integration with the machine}

\subsubsection{Final focus magnets}

The interaction region of ILD is designed to fulfil at the same time the requirements from the ILC machine as well as the needs of the detector. As the allowed focal length range of the inner final focus quadrupoles (QD0) for ILC ($3.5$~m~$\leq L^* \leq 4.5$~m) is smaller than the detector size, the QD0 magnets of the final lenses need to be supported by the detector itself. As a consequence, SiD and ILD will have their own pair of QD0 magnets that move together with the detector during push-pull operations. In contrast, the QF1 magnets of the final lenses with a focal length of $L^*=9.5$~m are not supported by the detectors and stay on the beam line during detector movements. A set of vacuum valves between the QD0 and the QF1 magnets define the break point for the push-pull operations. The biggest concerns for the QD0 support systems are the alignment and the protection against ground motion vibrations. The limit on the vibration amplitudes is given at 50~nm within the 1~ms long ILC bunch train~\cite{Parker:2009zz}.

\begin{figure}[t]
\centerline{\includegraphics[width=0.65\columnwidth]{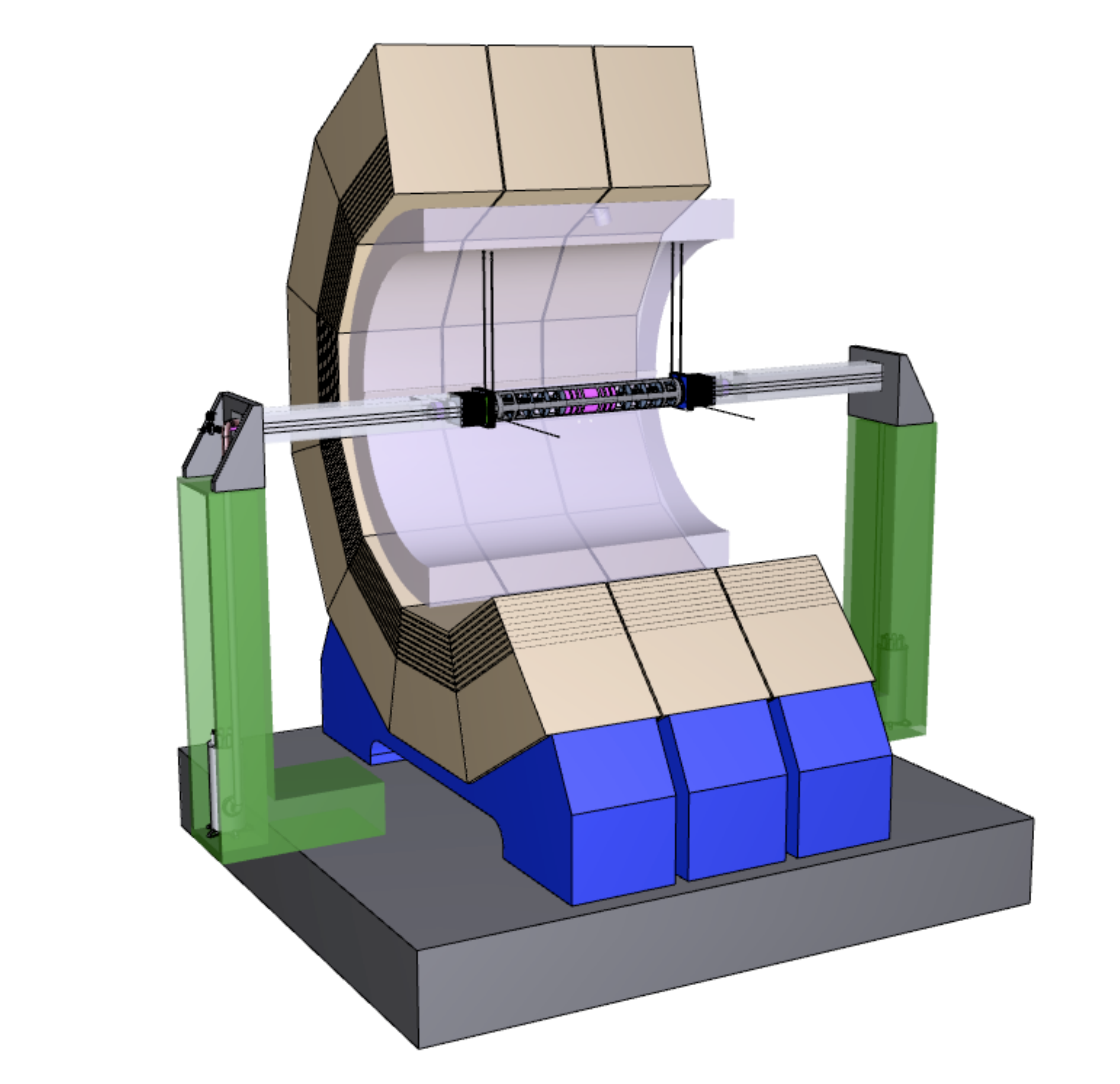}}
\caption{Support system of the QD0 magnets in ILD. The inner parts of the detector and the end caps are not shown~\cite{Group:2010eu}.}\label{Fig:qd0support}
\end{figure}

Due to these tight requirements, the support of the magnets in the detector is of special importance. ILD has chosen a design where the magnets are supported from pillars that are standing directly on the transport platform. In the detector, the magnets are supported by a system of tie rods from the cryostat of the solenoid coil. This design de-couples the detector end caps from the QD0 magnets and allows a limited opening of the end caps also in the beam position without the need to break the machine vacuum~(c.f.\ figure~\ref{Fig:ILD_opening}). In addition, the QD0 magnets are coupled via the pillar directly to the platform and limit in that way the number of other vibration sources. 
Simulations taking into account realistic ground motion spectra for different sample sites have been done to understand the vibration amplification in the QD0 support system~\cite{Yamaoka:2010}. These studies show, that with the exception of very noisy sites, the requirements for the QD0 magnets are fulfilled with large safety margins. Even if the additional amplification characteristics of the platform (c.f.\ \cite{Oriunno:2011}) are taken into account, the total integrated vibration amplitudes are in the order of less than 10~nm for frequencies above 5~Hz.

Also the proper alignment of the QD0 magnets with respect to the axis that is defined by the QF1 quadrupoles is of crucial importance. While the alignment accuracy of the detector axis after the movement into the beamline is moderate (horizontal: $\pm$~1~mm and $\pm$~100~$\mu$rad), the requirements for the initial alignment of the quadrupoles are much tighter: $\pm$~50~$\mu$m and $\pm$~20~$\mu$rad. An alignment system that comprises an independent mover system for the magnets and frequency scanning interferometers is part of the detector design.

\subsubsection{Interaction region}

\begin{figure}[t]
\centerline{\includegraphics[width=0.95\columnwidth]{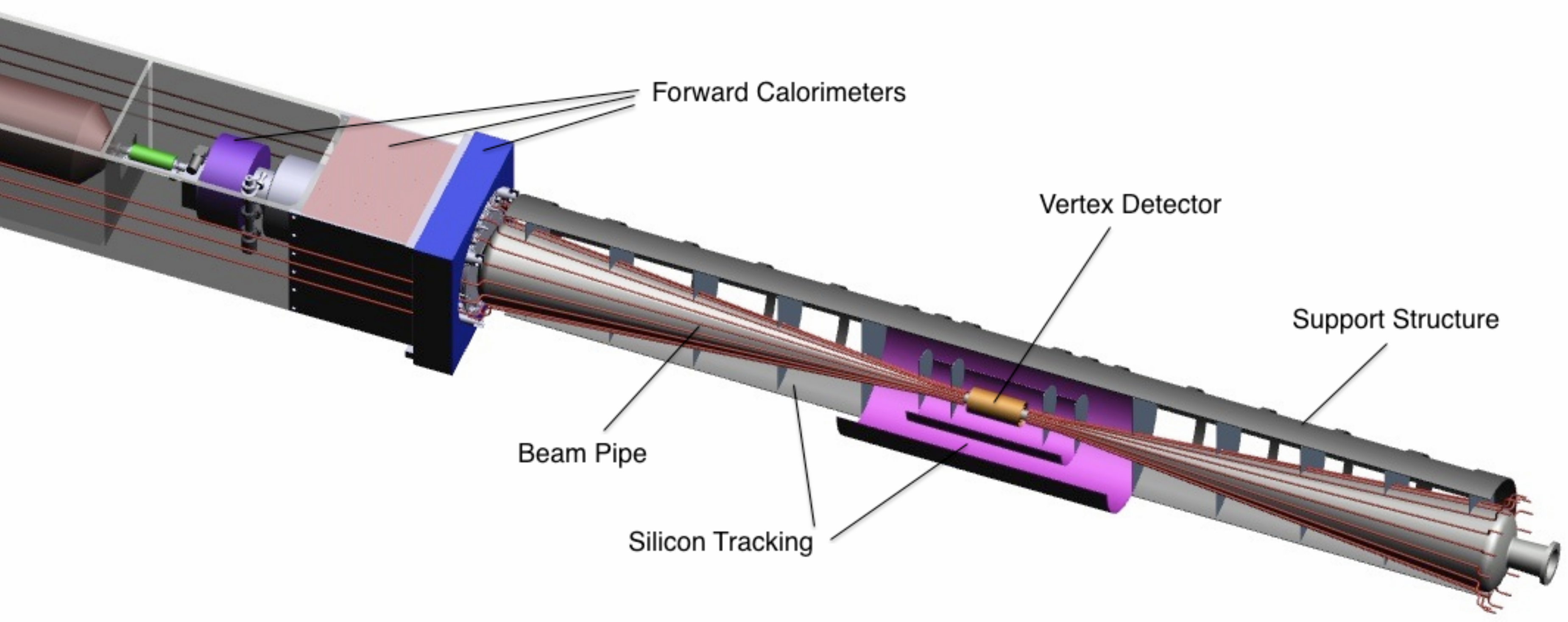}}

\caption{The interaction region of ILD~\cite{Group:2010eu}.}\label{Fig:interaction_region}
\end{figure}

The central interaction region of ILD comprises the beam pipe, the surrounding silicon detectors, the forward calorimeters and the interface to the QD0 magnets (c.f.\ Figure~\ref{Fig:interaction_region}).

The most delicate part of this region is the very light beam pipe made from Beryllium, that is surrounded by the vertex detector and the intermediate silicon tracking devices. A carbon fibre reinforced cylindrical structure will form the mechanical support for these elements. This tube is attached to the inner field cage of the surrounding time projection chamber (not shown in the figure). As the horizontal alignment tolerance of the detector axis after push-pull operations is $\pm$1~mm, an adjustment system is needed to eventually re-align the tube structure with the beam pipe and the inner tracking detectors. This is especially important to keep the stay-clear distances to the tracks of the beam induced background particles within the beam pipe.
mT
The beam pipe opens conically away from the interaction point to allow enough space for the beam induced background, most importantly the electron-positron pairs from beamstrahlung. The shape of the beam pipe results in a rather large volume that needs to be kept evacuated by means of of vacuum pumps that are on both sides as far as 3.3~m away from the interaction point. Simulations show however, that the vacuum requirements for the ILC can be met~\cite{Group:2010eu}.

The forward calorimeters have a two-fold purpose. They enlarge the hermeticity of the detector for physics analyses, but they also serve as beam diagnostic devices by measuring the patterns from the beamstrahlung pairs~\cite{Group:2010eu}.

\subsection{Detector services}
\label{Sec:Services}
A number of service and supply equipments needs to be established for the running and the maintenance of ILD. The arrangement of the services depends on the technical requirements and can be sorted according to their proximity to the detector. Primary services should be located on the surface above the experimental hall. They comprise usually large and sometimes noisy facilities like water chillers, high voltage transformers, auxiliary power supplies (Diesel generators), Helium storage and compressors, and gas storage systems. Secondary services will be placed into the underground cavern in dedicated service areas. Examples are cooling water distributions, power supplies, gas mixture systems, power converters, and parts of the cryogenic system for the detector (He liquefier and re-heater, control system). As the detector will not be disconnected during the push-pull operations, all supplies that go directly to the detector will be run in flexible cable chains. As the supply with cryogenic Helium needs to be established also during the detector movement, flexible cryogenic lines are foreseen. The detector will carry those services on-board that need to stay close or directly at the detector. Examples are the He system for the QD0 magnets, on-board electronics and the electronic containers.

\section{Requirements for the experimental area}

\subsection{Underground hall design}
\begin{figure}[ht]
\centerline{\includegraphics[width=1.0\columnwidth]{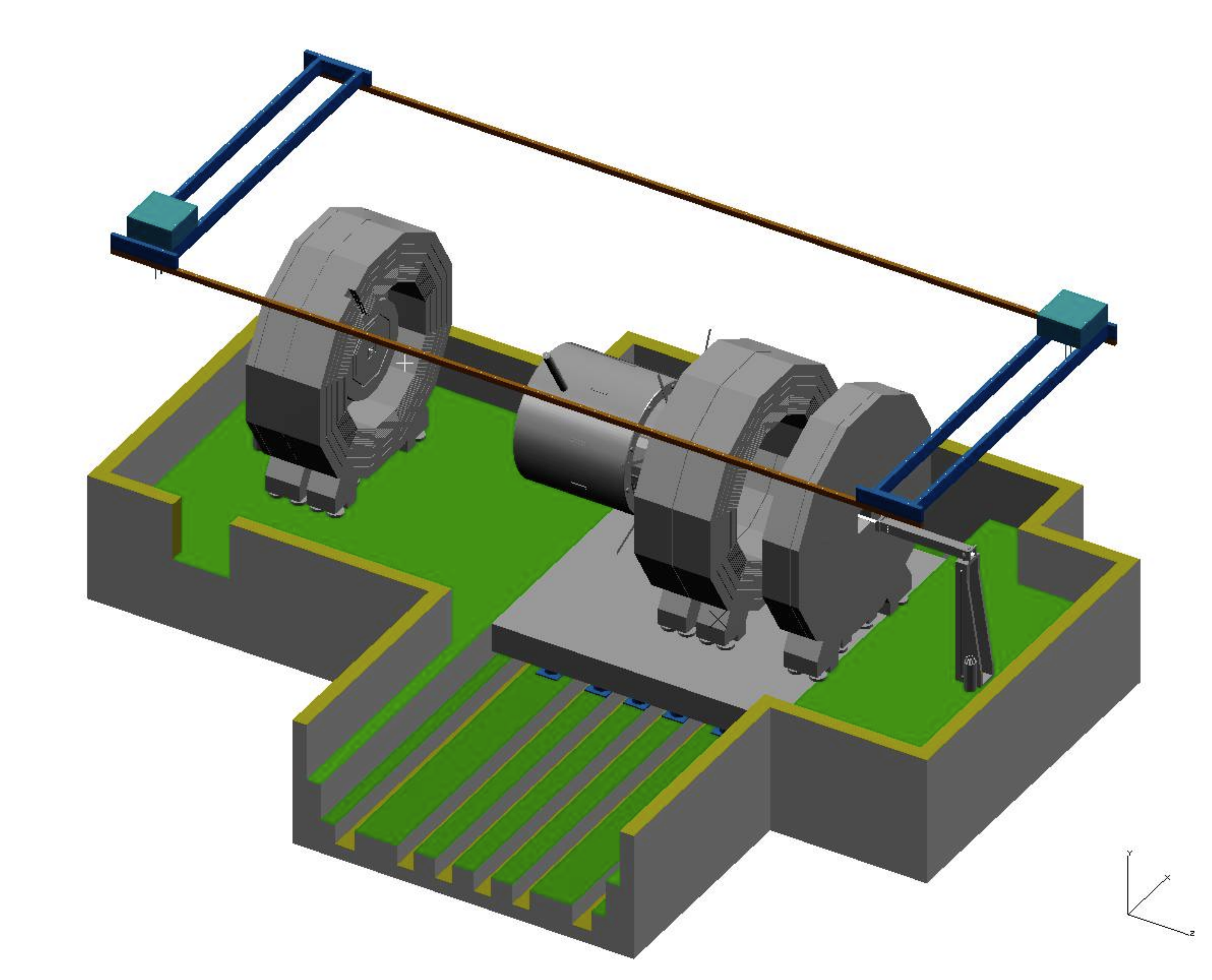}}

\caption{Conceptual design of the underground facilities for ILD. The detector is opened in the maintenance position, the crane coverage is shown~\cite{Sinram:2011}.}\label{Fig:ILD_underground}
\end{figure}

The discussions between both detector concepts and the civil facility experts of the ILC project are converging on an underground hall design that follows a z-shape floor layout as indicated in figure~\ref{Fig:Push-pull}~(left). The common interaction point is in the middle of the hall, the detectors move in and out of the beam position on their transport platforms. Alcoves in the maintenance positions allow for lateral space that is needed to open the detectors. Figure~\ref{Fig:ILD_underground} shows a recently developed design for the maintenance position for ILD.
The detector is shown in fully opened position that allows for the removal of the large detector parts. The biggest element that might need to be removed from the detector (though not in routine maintenance periods) is the superconducting solenoid. Enough space is foreseen to manoeuvre the parts of the detector in the hall and bring them safely to the vertical access shafts. In addition, space for the detector services (c.f.\ section~\ref{Sec:Services}) is available in this design.

The size and location of the vertical access shafts is still under study. The preferred solution is to foresee one central big shaft directly above the interaction point with a diameter of $\approx$~18~m. This shaft would be used during the assembly of both detectors where the big parts are pre-assembled on the surface and then lowered through the big shaft directly onto the respective transport platform. Two smaller diameter shafts ($\approx$~10~m) are needed in the maintenance positions to allow access from the surface while one detector is at the beam position and blocks the access to the big shaft. Additional smaller shafts for elevators and services might be needed as well. An overall optimisation of the layout and number of the shafts with respect to the functionality and the cost is under study.

As the yoke rings will be moved on air pads within the hall, the crane covering the maintenance area needs to have a modest capacity of preferably 2~$\times$~40~t. However, a temporary hoist with a capacity of up to 3500~t is needed on the surface over the main access shaft to lower the big detector parts during the primary assembly.\\

{\em Note:} the underground hall design discussed here would be chosen for the ILC sites that allow vertical access via relatively short shafts. This needs to be modified significantly for the Asian ILC reference sites (c.f.\ section~\ref{Sec:Japan}). A detailed design following these very different requirements is under study at this time.

\subsection{Shielding}
\subsubsection{Radiation}
\begin{figure}[t]
\centerline{\includegraphics[width=0.75\columnwidth]{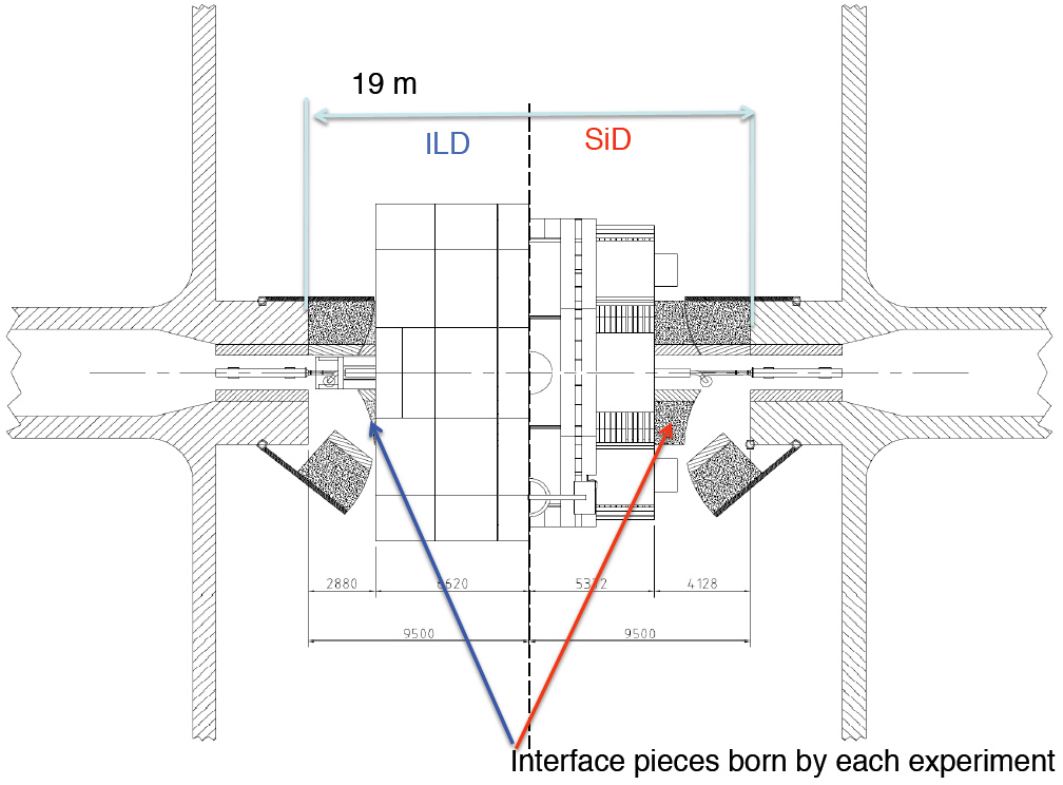}}
\caption{Design of the beamline shielding compatible with two detectors of different sizes~\cite{Elsen:2011zz}.}\label{Fig:shielding}
\end{figure}The ILD detector is self-shielding with respect to ionising radiation that stems from maximum credible beam loss scenarios~\cite{Sanami:2009}. Additional shielding in the hall is necessary to fill the gap between the detector and the wall in the beam position. The design of this beamline shielding needs to accommodate both detectors, SiD and ILD, that are of significant size differences. A common `pac-man' design has been developed, where the movable shielding parts are attached to the wall of the detector hall - respectively to the tunnel stubs of the collider - and match to interface pieces that are borne by the experiments (c.f.\ figure~\ref{Fig:shielding}).

\subsubsection{Magnetic fields}
\begin{figure}[t]
\centerline{\includegraphics[width=0.65\columnwidth]{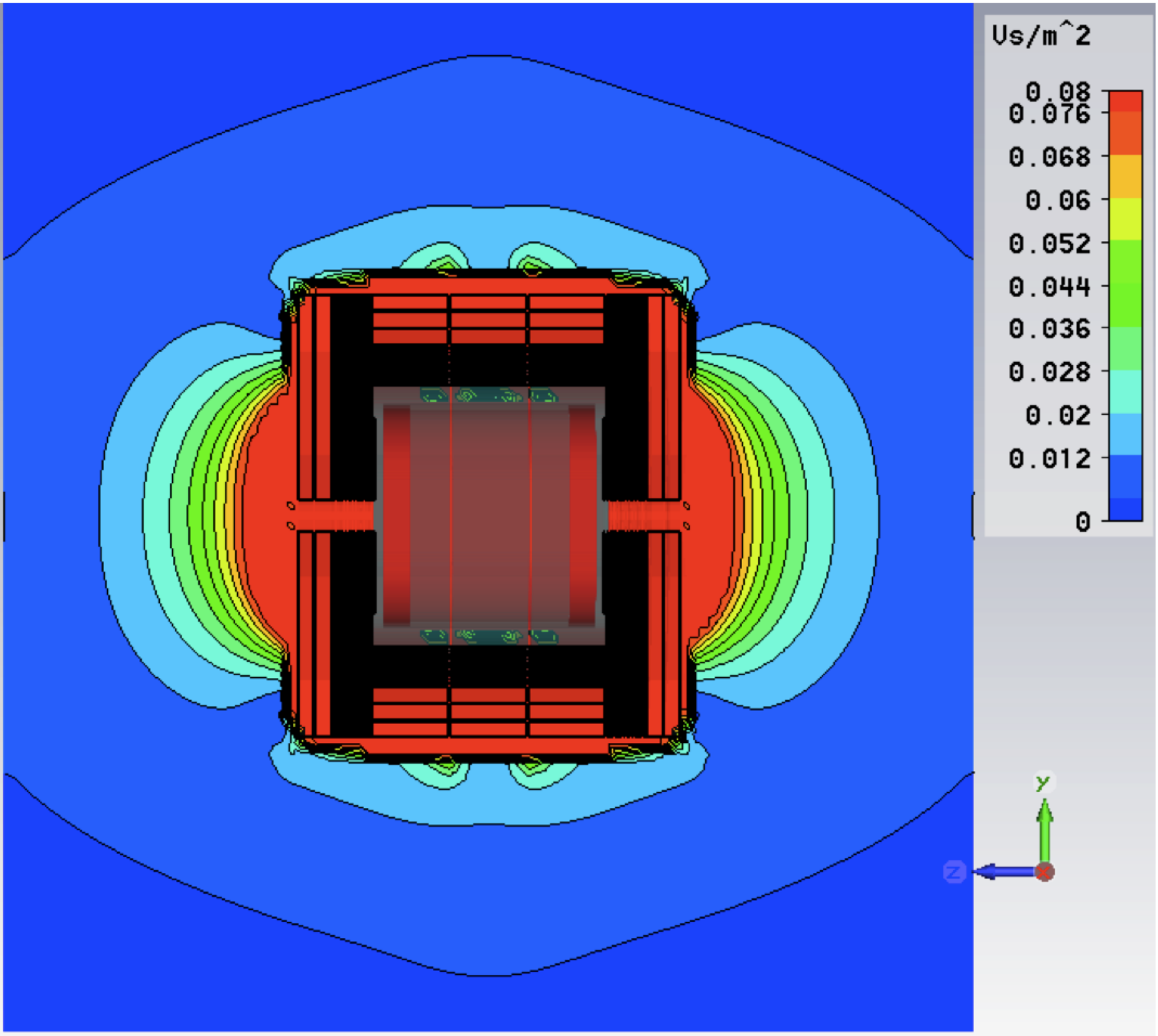}}
\caption{Magnetic stray fields from the detector solenoid~\cite{Group:2010eu}.}\label{Fig:strayfield}
\end{figure}

The magnetic stray fields outside the iron return yoke of the detector need to be small enough to not disturb the other detector during operation or maintenance. A limit for the magnetic fields has been set to 5~mT at a lateral distance of 15 m from the beam line~\cite{Parker:2009zz}. This allows the use of standard iron-based tools at the other detector. The design of the ILD return yoke has been tested carefully for the fringe fields. Figure~\ref{Fig:strayfield} shows the magnetic fields that have been simulated for a central solenoid field of 4~T.

\section{Summary and outlook}

Significant efforts in the worldwide ILC MDI work have been spent to develop an engineering design of the experimental environment for the two planned detectors in the push-pull scheme. While the conceptual design of all relevant infrastructures has been defined by now, the work is concentrating on the finalisation of the engineering specifications that will form the basis of the respective parts of the ILC Technical Design Report and the accompanying detector Detailed Baseline Descriptions that are envisaged to be published by the end of 2012.

\section{Acknowledgments}

The work on the ILC Machine-Detector Interface and the design of the detector related conventional facilities is a collaborative effort between the detector concepts and the respective ILC machine working groups. This report includes therefore the efforts of many people within the global ILC endeavour. I am especially grateful for the support of the members of the ILD MDI/Integration Group, the ILC MDI Common Task Group, the ILC Beam Delivery System Group, and the ILC Conventional Facilities Group.

Parts of this work were supported by the Commission of the European Communities, contract 206711 "ILC-HiGrade", and by the Helmholtz Association, contract HA-101 "Physics at the Terascale".





\begin{footnotesize}

\bibliographystyle{utphys_mod}
\bibliography{LCWS11_buesser}



\end{footnotesize}


\end{document}